\documentclass[notitlepage,aps,prl,reprint,twocolumn,longbibliography,superscriptaddress]
{revtex4-2}
\usepackage{graphicx}
\usepackage{amsmath}
\usepackage{amssymb}
\usepackage{comment}
\usepackage[colorlinks, allcolors=blue]{hyperref}
\usepackage[all]{hypcap}
\usepackage[mathlines]{lineno}
\usepackage{physics}
\usepackage{wrapfig}
\usepackage{lipsum}
\usepackage[normalem]{ulem}
\usepackage{siunitx}

\newcommand{\pref}[2]{\hyperref[#1]{\ref{#1}(#2)}}

\begin{document}
	\title{Collective dissipation engineering of interacting Rydberg atoms 
	}
	\author{Tao Chen}
\email{taochen@xjtu.edu.cn}
\affiliation{Department of Physics, University of Illinois at Urbana-Champaign, Urbana, IL 61801-3080, USA}
\affiliation{Department of Physics, The Pennsylvania State University, University Park, Pennsylvania 16802, USA}
\affiliation{School of Physics, Xi’an Jiaotong University, Xi’an 710049, China}
    \author{Chenxi Huang}
\affiliation{Department of Physics, University of Illinois at Urbana-Champaign, Urbana, IL 61801-3080, USA}
\affiliation{Department of Physics, The Pennsylvania State University, University Park, Pennsylvania 16802, USA}
    \author{Jacob P. Covey}
 \affiliation{Department of Physics, University of Illinois at Urbana-Champaign, Urbana, IL 61801-3080, USA}
	\author{Bryce Gadway}
\email{bgadway@psu.edu}
 \affiliation{Department of Physics, The Pennsylvania State University, University Park, Pennsylvania 16802, USA}
	\affiliation{Department of Physics, University of Illinois at Urbana-Champaign, Urbana, IL 61801-3080, USA}
	\date{\today}
 
\begin{abstract}

Engineered dissipation is emerging as an alternative tool for quantum state control, enabling high-fidelity preparation, transfer and stabilization, and access to novel phase transitions. We realize a tunable, state-resolved laser-induced loss channel for individual Rydberg atoms, in both non-interacting and strongly correlated settings. This capability allows us to reveal interaction-driven shifts of the exceptional point separating quantum Zeno and anti-Zeno regimes, and to demonstrate interaction-enhanced decay. By exploiting interaction-dependent energy level shifts, we observe a configuration-selective two-body Zeno effect that freezes target spin states. We theoretically show that when this mechanism is extended to many-body chains it allows for the dissipative distillation of unwanted spin configurations.
These experimental studies establish a versatile approach for exploring strongly interacting, open quantum spin systems, and opens possible new routines for dissipative preparation of correlated quantum states in Rydberg atom arrays.

\end{abstract}

\maketitle

Dissipative processes, such as the irreversible coupling of a system to its environment, generally occur alongside the flow of quantum information through measurements or projections. Uncontrolled dissipation typically introduces noise and degrades quantum coherences, thereby limiting the fidelity of quantum gates and quantum state transformations. 
Dissipation also presents new opportunities, and
one growing area of research involves dissipation engineering~\cite{Poyatos1996,Harrington2022} for applications in high-level quantum state manipulation~\cite{Bernu2008,Diehl2008,Verstraete2009,Morigi2015,Barontini2015}, readout~\cite{Zhu2025}, and stabilization~\cite{Kimchi2016,Lu2017,Leghtas2015,Ma2019,Ritter2025}. For example, recent experiments have demonstrated quantum error correction protocols leveraging the leakage channels of atomic qubits~\cite{Ma2023,Scholl2023}, demonstrating that loss can be not only mitigated, but used to improve the performance of quantum systems. In the non-interacting regime, tuning the dissipation across an exceptional point of an open quantum system, i.e., the coupling strength to the environment or an auxiliary reservoir, leads to a parity-time ($\mathcal{PT}$) symmetry breaking phase transition~\cite{El2018,Li2019,Naghiloo2019,Chen2022sc,Ding2021,Shi2025}, accompanied by a quantum anti-Zeno to Zeno crossover~\cite{Sun2023, Chen2021, Li2023, Zhou2023}. The quantum Zeno effect has been used for preparing non-classical state~\cite{Morigi2015,Signoles2014}, stabilizing quantum entanglement~\cite{Liu2016,Lingenfelter2024}, and generating decoherence-free subspaces~\cite{Facchi2002,Bretheau2015,Touzard2018,Bigorda2025}. 

Correlations between particles fundamentally reshape dissipation dynamics, producing a far richer but more complicated landscape than in the non-interacting scenario \cite{Lee2012,Lee2014, Marcuzzi2014,McDonald2022}. Many-body interactions can lead to the acceleration of dissipation rates~\cite{Wang2024Nexus} and the emergence of unstable dynamics~\cite{Koppenhofer2022,Pocklington2023}. 
The introduction of dissipation has also been recently demonstrated in superfluid junction experiments~\cite{Talebi2024,Huang2024} to enhance spin relaxation when strong two-body loss channels operate in the anti-Zeno regime. 
Conversely, strong dissipation can actually inhibit or suppress inelastic collisional loss via the continuous quantum Zeno effect, preventing interacting particles from populating lossy two-body configurations~\cite{Syassen2008,Zhu2014,Tomita2017} and with prospects for stabilizing exotic quantum states~\cite{zoller-stab,NH-FQH}.
While such many-body dissipative phenomena have been widely studied at the macroscopic level~\cite{Mendoza2016,Lu2021,Meng2022, Xie2024,Li2024,Zhao2025hk,liu2025,Zhang2025}, the microscopic description on how the two-body collisions (interactions) modify the thermalization and loss behaviors remains to be elucidated.

\begin{figure}[h]
	\includegraphics[width=0.48\textwidth]{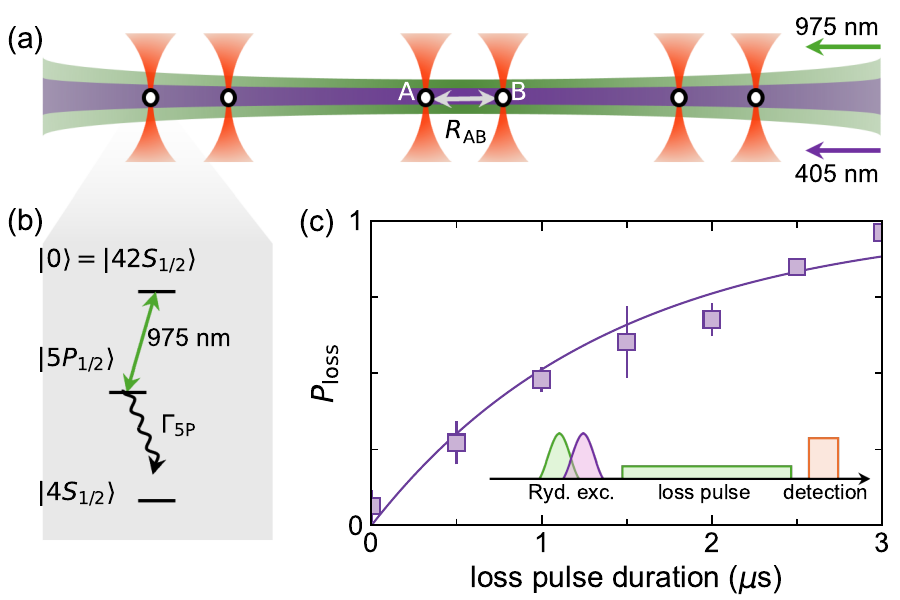}
	\caption{\textbf{Laser induced dissipation for individual Rydberg atoms.}
        \textbf{(a)}~Dimerized Rydberg atom arrays prepared in optical tweezers, with the 405 nm (purple) and 975 nm (green) two-photon excitation beams. The state-dependent dipolar interaction between atoms $A$ and $B$ is tuned by changing the interatomic distance $R_{\rm AB}$.
        \textbf{(b)}~Light induced dissipation for the Rydberg state $\ket{0}=\ket{42S_{1/2}}$. A weak 975 nm laser pulse is applied to couple the $\ket{0}\leftrightarrow \ket{5P_{1/2}}$ transition, followed by a rapid spontaneous decay to the ground state with the rate $\Gamma_{5P}/h\sim 1~{\rm MHz}$. 
        \textbf{(c)}~Measurement of the effective loss rate. Each individual atom is first prepared in $42S_{1/2}$ via two-photon Rydberg excitation, and then subjected to the weak resonant dissipative 975 nm light. Population lost to the ground state is detected via fluorescence imaging (see the inset time sequence). An exponential fit of the normalized loss versus the 975 nm laser pulse duration yields an effective loss rate $\gamma/h = 0.11(2)~{\rm MHz}$.
        Error bars indicate the standard errors from multiple independent measurements.
}
\label{fig1}
\end{figure}  

Rydberg atom arrays serve as a clean platform with flexible tunability for studying the interplay between strong correlations and dissipation, particularly in the few-body regime. Many theory proposals and potential applications have been shown for dissipative long-range spin chain models, e.g., observing Yang-Lee singularities in an Ising model with an imaginary transverse field \cite{Jian2021,Matsumoto2022,Shen2023,Xu2025}. However, the experimental realization with Rydberg atom arrays is still rarely investigated, 
especially on how to introduce a controllable loss channel \cite{Begoc2025} instead of solely relying on the intrinsic spontaneous decay of Rydberg states \cite{Marcuzzi2014,Landa2020,Shen2023,Xie2024}. Here, we demonstrate: (i) laser-induced tunable dissipation for individual Rydberg atoms; (ii) dipolar exchange interaction-driven $\mathcal{PT}$ symmetry breaking and shifting of exceptional points; and (iii) the protection of selected two-body spin configuration via the interaction-enabled quantum Zeno effect, which can be further used for state preparation or spin purification. These studies help to establish Rydberg atom arrays as a system poised to explore new frontiers of many-body open quantum systems.

\textit{Laser-induced controllable dissipation.--} 
Our experiment begins with a one-dimensional optical tweezer array of individual $^{39}$K atoms prepared in the internal ground state $\ket{g}$, followed by a two-photon Rydberg excitation to the $\ket{42S_{1/2}}$ state \cite{Chen2024a,Chen2024b,Chen2025}. As shown in Fig.~\ref{fig1}, we use a resonant 975 nm laser to weakly couple the transition $\ket{42S_{1/2}}\leftrightarrow \ket{5P_{1/2}}$. Due to the rapid spontaneous decay of the $\ket{5P_{1/2}}$ state, this coupling effectively introduces a controlled and tunable dissipation channel from the Rydberg manifold back to the ground state. The induced loss is characterized by monitoring the population in $\ket{g}$ as a function of the loss pulse duration, following the time sequence shown in Fig.~\pref{fig1}{c}. 
The measured population loss is renormalized by considering state preparation and measurement infidelity~\cite{SuppMats}.
The loss rate $\gamma$ typically stays on the order of $h\times 0.1~{\rm MHz}$, and can be continuously tuned by varying the 975 nm laser intensity. 

To clearly illustrate the influence of interactions on the dissipation, we perform post-selections to contrast the single-atom dynamics to those for atom pairs, similar to our previous work \cite{Chen2024a}. The strength of dipolar exchange can be precisely tuned by adjusting the interatomic spacing with optical tweezers. Our approach thus establishes a versatile and well-controlled platform to study open quantum few-body systems. Building upon this controllable dissipation mechanism, in the following we focus on how the presence of dipolar exchange interactions can enable emergent phenomena such as $\mathcal{PT}$ symmetry breaking and shifts of the system's exceptional points.

\begin{figure*}[t]
	\includegraphics[width=0.9\textwidth]{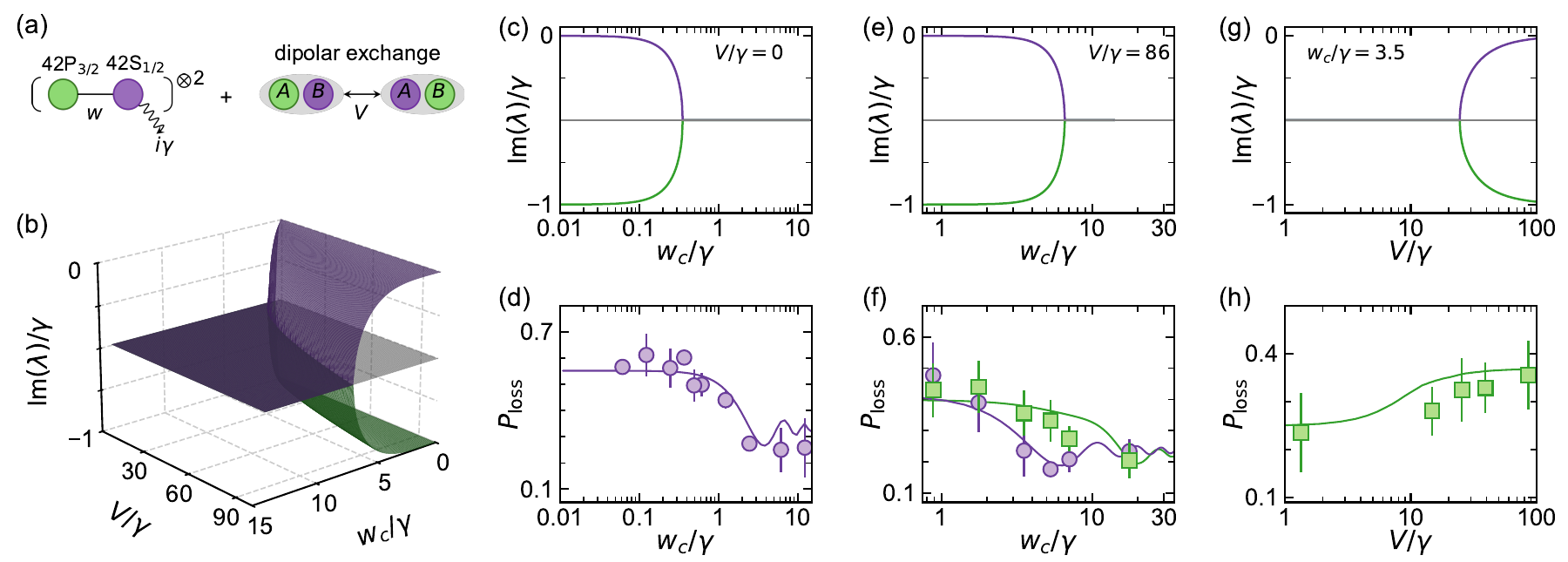}
	\caption{\textbf{Interaction-enabled phase transitions and dissipation enhancement.}
        \textbf{(a)}~Experimental realization of the interacting two-body dissipative spin model. We encode $\ket{1}=\ket{42P_{3/2},m_J=1/2}$ and $\ket{0}=\ket{42S_{1/2}, m_J=1/2}$, respectively, and they are resonantly coupled by microwaves with a strength $w$. The $\ket{42S_{1/2}}$ state has a laser-induced effective loss term $\gamma$. The atom pair (labeled $A$ and $B$) also experiences dipolar exchange with a strength $V=C_3/(2R_{\rm AB}^3)$, where $C_3=h\times 1502~{\rm MHz}~\mu{\rm m}^3$ and $R_{\rm AB}$ the interatomic distance. 
        \textbf{(b)}~Imaginary parts of the eigenenergy values $\lambda$ of the effective two-body non-Hermitian Hamiltonian (\ref{eq2}) under different $V/\gamma$ and $w_c/\gamma$ ratios (with $w_c=\sqrt{2}w$). \textbf{(c)}~Im$(\lambda)$ vs. the $w_c/\gamma$ ratio for the non-interacting case ($V=0$). \textbf{(d)}~Normalized measured population loss after $1~\mu{\rm s}$ time evolution for different $w_c/\gamma$ ratios with $V=0$. Here, the two atoms are initialized in the $\ket{00}$ state, and $\gamma/h=0.13(2)$ MHz. \textbf{(e)}~Im$(\lambda)$ vs. the ratio $w_c/\gamma$ under strong dipolar interactions with $V/\gamma=86$. \textbf{(f)}~Normalized measured population loss after $1~\mu{\rm s}$ time evolution for different $w_c/\gamma$ ratios (green squares), with $V/h=6.88(16)~{\rm MHz}$ and $\gamma/h=0.08(2)~{\rm MHz}$. For comparison, the results for the non-interacting case are shown (purple circles) for the same parameters. \textbf{(g)}~Im$(\lambda)$ vs. the ratio $V/\gamma$ with $w_c/\gamma=3.5$. \textbf{(h)}~Normalized measured population loss after $1~\mu{\rm s}$ evolution under different interaction strengths with $w/h=0.20(1)~{\rm MHz}$ and $\gamma/h=0.08(2)~{\rm MHz}$.
        In (d,f,h), solid lines indicate the simulation results from the Lindblad equation (\ref{eq1}) with state preparation and measurement infidelity included. 
        Error bars are standard errors from multiple independently measured data sets.
}
\label{fig2}
\end{figure*}  

\textit{Interplay between dissipation and interactions.--}
We first investigate the dissipative behaviors of a lossy two-level system under dipolar exchange interactions. We introduce a second Rydberg state $\ket{1}=\ket{42P_{3/2},m_J=1/2}$ that is coherently coupled with microwave to the lossy $\ket{0}=\ket{42S_{1/2},m_J=1/2}$ state, as shown in Fig.~\pref{fig2}{a}. For an atom pair (labeled $A$ and $B$), the system is described by the Lindblad master equation
\begin{equation}\label{eq1}
 \frac{\partial \rho}{\partial t} = -i[H_{\rm tot}, \rho] + L_0\rho L_0^\dagger -\frac{1}{2}\{L_0^\dagger L_0, \rho\}
\end{equation}
where $\rho$ is the density matrix, and $H_{\rm tot} = w(\ket{0}_A\bra{1} + \ket{0}_B\bra{1} ) + V\ket{0}_A\bra{1}\otimes\ket{1}_B\bra{0} + {\rm H.c.}$ with $w$ and $V$ denoting the coupling strength and the dipolar exchange strength, respectively. The collapse operator $L_0 = \sqrt{\gamma}(\ket{g}_A\bra{0}\otimes I_B + I_A\otimes\ket{g}_B\bra{0}$ with $\gamma$ the laser-induced effective loss rate and $I$ the identity. By eliminating the quantum jumps associated with decay from the $\ket{0}$ state \cite{Chen2021sc}, the dynamics within the $\{\ket{0}, \ket{1}\}$ manifold are governed by a non-Hermitian Hamiltonian. Under the basis $\{\ket{00}, \ket{+_{}}, \ket{11}\}$ where $\ket{+_{}}=(\ket{01}+\ket{10})/\sqrt{2}$ (with the antisymmetric state $\ket{-_{}}=(\ket{01}-\ket{10})/\sqrt{2}$ fully decoupled), the effective Hamiltonian reads
\begin{equation}\label{eq2}
 H = \left(\begin{matrix}-i\gamma & w_c & 0 \\ w_c & V-i\gamma/2 & w_c \\ 0 & w_c & 0\end{matrix}\right) \ ,
\end{equation}
with the collective microwave coupling strength $w_c = \sqrt{2}w$. Note that we neglect the spontaneous decay of the $42P$ state as its rate is generally at least an order of magnitude smaller than the rate of induced dissipation. 

Figure \pref{fig2}{b} presents the imaginary parts of the eigenvalues $\lambda$ of the Hamiltonian (\ref{eq2}) under different microwave coupling and interaction strengths. Varying either $w_c$ or $V$ shifts the location of the exceptional points, which generally delineate the boundary between the quantum Zeno and anti-Zeno regions \cite{Li2023,Chen2021,Sun2023}. We therefore expect to observe different dissipation behaviors when scanning across the exceptional points. In the experiment, we initialize atoms $A$ and $B$ both in the $\ket{0}$ state, and then simultaneously turn on both microwave coupling and the dissipation laser. We characterize dissipation by measuring the population loss out of the $\{\ket{0}, \ket{1}\}$ subspace after $1~\mu{\rm s}$ evolution time, providing a qualitative indication to the different behaviors across the $\mathcal{PT}$ symmetry breaking transition. 

We first examine how the dipolar exchange interaction modifies the dissipation dynamics. For the non-interacting case ($V=0$), the exceptional point locates at $w_c/\gamma=\sqrt{2}/4$, consistent with $w/\gamma=1/4$ from previous studies~\cite{Li2019,Naghiloo2019}; see Fig.~\pref{fig2}{c}. In experiment, by scanning the coupling strength, we observe suppressed population loss for large $w_c/\gamma>\sqrt{2}/4$, as shown in Fig.~\pref{fig2}{d}. This indicates that the system resides in the $\mathcal{PT}$ symmetric regime where the quantum Zeno effect emerges and hinders the dissipation process 
as a strong $w_c$ makes the $\ket{0} \leftrightarrow \ket{1}$ oscillation dominate the dynamics~\cite{Facchi2002, Signoles2014} 
(see the Supplement for time evolution dynamics in different regimes~\cite{SuppMats}). 
In the $w_c \to \infty$ limit, the $\ket{0}$ state is effectively decoupled from the resonant 975~nm dissipation channel, leading to negligible loss. 
However, introducing a strong interaction $V=86\gamma$ significantly moves the exceptional point to $w_c/\gamma\sim 7$ position; see Fig.~\pref{fig2}{e}. Under this condition, for small $w_c/\gamma$ ratios within the $\mathcal{PT}$ symmetry broken regime, the system experiences substantially faster dissipation speed compared to the non-interacting case which already lies in the $\mathcal{PT}$ symmetric regime. Figure \pref{fig2}{f} contrasts the loss behaviors for a single atom and an interacting atom pair, demonstrating that the interaction can enhance the dissipation before $w_c/\gamma$ crosses the exceptional point. Due to such a shift, the Zeno-suppressed loss occurs at relatively larger $w_c/\gamma$ values compared to the non-interacting case. 

To further identify the interaction-enhancement of dissipation, we investigate the shift of the exceptional point at fixed $w_c$. Figure \pref{fig2}{g} depicts a clear interaction-induced $\mathcal{PT}$ symmetry breaking phase transition. Accordingly, in Fig.~\pref{fig2}{h}, increasing the interaction strength in our experiment leads to increased atom loss, providing direct evidence of interaction-enhanced two-body
dissipation. While a similar phenomenon has been observed in cold atomic systems under mean-field contact interactions~\cite{Wang2024Nexus}, our experiment shows that dipolar exchange can be harnessed to tune dissipative phase transitions in systems with microscopic control.
Our observations not only support phase transition across the exceptional point from Hamiltonian~(\ref{eq2}), but also agree with predictions based on a full Lindblad master equation~(\ref{eq1}).

\begin{figure}[]
	\includegraphics[width=0.9\columnwidth]{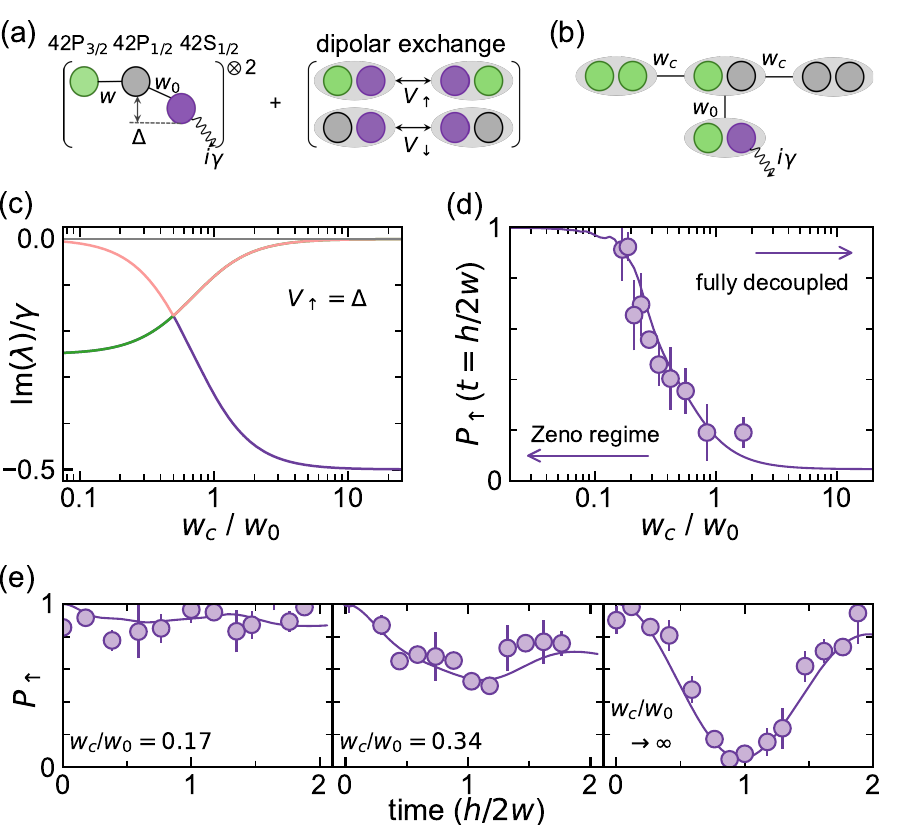}
	\caption{\textbf{Dissipation engineering for selective configuration loss in Rydberg atom pairs.}
        \textbf{(a)}~Experimental scheme. We encode $\ket{\uparrow}=\ket{42P_{3/2},m_J=1/2}$ and $\ket{\downarrow}=\ket{42P_{1/2},m_J=1/2}$ which are coupled by two-photon Raman transition \cite{SuppMats}. The $\ket{\downarrow}$ is coupled by another microwave tone to $\ket{0}=\ket{42S_{1/2},m_J=1/2}$, with a coupling strength $w_0$ and detuning $\Delta$. The $42S_{1/2}$ experiences light-induced dissipation with an effective rate $\gamma$; see Fig.~\pref{fig1}{b}. The three Rydberg states undertake two dipolar exchange processes $\ket{\uparrow 0}\leftrightarrow\ket{0 \uparrow}$ and $\ket{\downarrow 0}\leftrightarrow\ket{0 \downarrow}$ with the strength of $V_\uparrow$ and $V_\downarrow$ respectively.
        \textbf{(b)}~Reduced configuration in pair basis under the condition $|V_\uparrow-V_\downarrow|\gg w, \gamma$, $|\Delta-V_\downarrow|\gg w_0, \gamma$. The loss channel is activated by letting $\Delta=V_{\uparrow}$.
        \textbf{(c)}~The imaginary parts of the eigenenergy values $\lambda$ for the Hamiltonian (\ref{eq3}) for different $w_c/w_0$ ratios with $V_\uparrow=\Delta$.
        \textbf{(d)}~By letting $\Delta=V_{\uparrow}=h\times4.0(1)~{\rm MHz}$ and starting from $\ket{\uparrow\uparrow}$, we plot the normalized population in $\ket{\uparrow}$ state, $P_{\uparrow}$ at $t=h/2w (\sim 1.7~\mu{\rm s})$, as a function of the effective hopping ratio $w_c/w_0$. Here we use $\gamma/h=0.16(2)~{\rm MHz}$ and $w/h=0.15(2) ~{\rm MHz}$ and vary $w_0$. For large $w_0$, the system lies in Zeno regime, atoms in $\ket{\uparrow\uparrow}$ are protected from loss. For $w_0\to 0$, the atom pair oscillates between $\ket{\uparrow\uparrow}$ and $\ket{\downarrow\downarrow}$ and consequently at $\pi$ time it populates state $\ket{\downarrow\downarrow}$ with $P_\uparrow=0$. 
        Solid line is numerical simulation by solving the Lindblad equation with all relevant parameters and their uncertainties. The state preparation efficiency and fluctuations of the interaction strength stemming from finite temperature effects are also included. 
        \textbf{(e)}~Time evolution of $P_\uparrow$ for 3 different $w_c/w_0$ ratios: 0.17 (left), 0.34 (middle) and $\infty$ (right), with $w_0/h=\{1.25(2), 0.63(2), 0\}$~MHz, respectively. Other parameters are the same as in (d). 
        All error bars are standard errors from multiple measured data sets.
}
\label{fig3}
\end{figure}

\textit{Selective dissipation of specific spin configurations.--}
The ability to engineer state-dependent dissipation offers a powerful resource for quantum state control. To make connection to previous studies of dissipative stabilization in reactive molecular gases~\cite{Syassen2008,Zhu2014}, we further explore if it is possible to engineer loss channels to selectively protect certain Rydberg pair state configurations while rapidly depleting other undesired states. In experiment, we encode a spin-1/2 system with $\ket{\uparrow}=\ket{42P_{3/2},m_J=1/2}$ and $\ket{\downarrow}=\ket{42P_{1/2},m_J=1/2}$, which are coupled by a two-photon Raman transition with microwave driving via the upper $\ket{43S_{1/2},m_J=-1/2}$ state \cite{SuppMats}. These pseudospin states, having the same parity, do not play host to dipolar interactions. 
Still, we can use the interactions between these states and auxiliary states to introduce correlated loss from this non-interacting state manifold. As shown in Fig.~\pref{fig3}{a}, $\ket{\downarrow}$ is connected to the dissipative state $\ket{0}=\ket{42S_{1/2},m_J=1/2}$ via another microwave tone with a coupling strength $w_0$ and a detuning $\Delta$. For an atom pair, this auxiliary $42S$ hosts dipolar exchange interactions with both of the $42P$ states, with exchange strengths of $V_\uparrow$ and $V_\downarrow$, respectively.
For our chosen states, $V_\uparrow\approx2V_\downarrow$ as the coefficients $\{C_3^\uparrow, C_3^\downarrow\}/h=\{1502, 756\}~{\rm MHz~\mu m^3}$. This system again can be described by the Lindblad master equation (\ref{eq1}), now for a $16\times 16$ density matrix $\rho$, with the Hamiltonian $H_{\rm tot}=\sum_{\alpha=A,B}(w\ket{\uparrow}_\alpha \bra{\downarrow} + w_0\ket{0}_\alpha\bra{\downarrow} - \Delta\ket{0}_\alpha\bra{0} ) + V_\uparrow\ket{\uparrow}_A\bra{0}\otimes\ket{0}_B\bra{\uparrow} + V_\downarrow \ket{\downarrow}_A\bra{0}\otimes\ket{0}_B\bra{\downarrow}+ {\rm H.c.}$. By tuning the loss channel parameters, $w_0$ and $\Delta$, we can map the dynamics onto different dissipative spin models. Under the condition $|V_\uparrow-V_\downarrow|\gg w, \gamma$ and $|\Delta-V_\downarrow|\gg w_0, \gamma$, the dynamics are effectively governed by an effective non-Hermitian Hamiltonian \cite{SuppMats} 
\begin{equation}\label{eq3}
 H_{\rm eff} = \left(\begin{matrix}0 & w_c & 0 & 0 \\ w_c & 0 & w_c & w_0 \\ 0 & w_c & 0 & 0 \\ 0 & w_0 & 0 & V_\uparrow - \Delta - i\gamma/2\end{matrix}\right) \ ,
\end{equation}
with coupling $w_c=\sqrt{2}w$ in the 
$\{\ket{\uparrow\uparrow}, \ket{+}_{\uparrow\downarrow}, \ket{\downarrow\downarrow}, \ket{+}_{\uparrow 0}\}$
pair spin basis, with triplet $\ket{+}_{rs}=(\ket{rs} + \ket{sr})/\sqrt{2}$. 

Figure \pref{fig3}{b} illustrates the simplified model by letting $V_\uparrow = \Delta$. For a fixed dissipation rate $\gamma$, the spin dynamics show different behaviors when tuning the $w_c/w_0$ ratio. We focus on the two extreme limits. 
First, when $w_c \gg w_0$ (i.e., small $w_0$), the lossy state $\ket{0}$ is fully decoupled from the spin-1/2 manifold (here the dissipation is dominated by the eigenstate $\ket{\lambda}=\ket{+}_{\uparrow 0}$ that takes an imaginary eigenvalue of $-i\gamma/2$; see Fig.~\pref{fig3}{c}).
Starting from $\ket{\uparrow\uparrow}$, the system undergoes coherent Rabi oscillations, 
reaching $\ket{\downarrow\downarrow}$ at the $\pi$ time $t=h/2w$, as shown in the right panel in Fig.~\pref{fig3}{e}.
Figure~\pref{fig3}{d}, which shows the $\pi$-time population in $\ket{\uparrow}$ versus $w_c / w_0$, displays this regime of the decoupled lossy channel at the right.

In the opposing limit $w_c \ll w_0$, strong dissipation process happens primarily within the subspace $\{\ket{+}_{\uparrow\downarrow}, \ket{+}_{\uparrow 0}\}$. Here, a large $w_0$ can be considered a strong projective measurement over the $\ket{+}_{\uparrow\downarrow}$, inducing the quantum Zeno effect and giving rise to a protected Zeno subspace $\{\ket{\uparrow\uparrow}, \ket{\downarrow\downarrow}\}$. In this two-body Zeno regime, pairs initialized in $\ket{\uparrow\uparrow}$ experience inhibited spin-flipping, becoming effectively trapped in their spin configuration by correlated loss. The dynamics in this limit are shown in the left panel of Fig.~\pref{fig3}{e}, relating to the left region of Fig.~\pref{fig3}{d} with a large $\pi$-time $P_\uparrow$ population.

The clearest version of this  
two-body Zeno process occurs when $\Delta\gg w_0\gg w_c$, starting from $\ket{\uparrow\uparrow}$. Here, the dynamics of a single atom and a pair differ qualitatively~\cite{SuppMats}. For a single atom, the large detuning $\Delta$ effectively isolates the $\ket{0}$ state from the microwave-driven $\ket{\uparrow}\leftrightarrow\ket{\downarrow}$ transition. The atom therefore undergoes conventional Rabi oscillations, 
unaffected by dissipation. By contrast, for two atoms, the large $w_0$ enforces strong virtual transitions into the lossy channel, which freezes the pair entirely in the $\ket{\uparrow\uparrow}$ state via the quantum Zeno effect. Such a pair-level Zeno protection is thus a genuinely collective phenomenon: it does not manifest in the dynamics of single particles and arises only from the interplay between the strong dissipation and dipolar exchange. 

To further understand these selectively lossy phenomena, we perform another approximation by treating the lossy subspace $\{\ket{+}_{\uparrow\downarrow}, \ket{+}_{\uparrow 0}\}$ as a direct loss channel from the $\ket{+}_{\uparrow\downarrow}$ state with an effective rate $\gamma_{\rm eff}\approx 4w_0^2/\gamma$ \cite{Sun2023, Li2023}. While this stationary form is valid for $w_0\ll\gamma$, it provides a qualitatively accurate description of the experimental results in quantum Zeno regime with $w_0>\gamma$ \cite{Lapp2019,Gou2020}. Under this approximation, we can rewrite the two-atom Hamiltonian (\ref{eq3}) as 
\begin{equation}\label{eq4}
 H_{\rm eff} = w(\sigma_A^x + \sigma_B^x) - i\gamma_{\rm eff}\ket{+}_{\uparrow\downarrow}\bra{+}
\end{equation}
with $\sigma_x$ denoting the Pauli matrix. This simplified Hamiltonian offers an explicit picture of the competition between coherent spin flips $w$ and the effective loss rate $\gamma_{\rm eff}$. As illustrated in Figs.~\pref{fig3}{d,e}, increasing $\gamma_{\rm eff}$ (i.e., increasing $w_0$) makes the atom pair tend to avoid populating the lossy $\ket{+}_{\uparrow\downarrow}$ state. This suppression marks a crossover from the regime of independent single-atom dynamics, where dissipation plays no role, to the fully protected two-body Zeno regime, where the atom pair remains frozen in its initial configuration and spin-flip processes are effectively inhibited. 

\begin{figure}[]
	\includegraphics[width=\columnwidth]{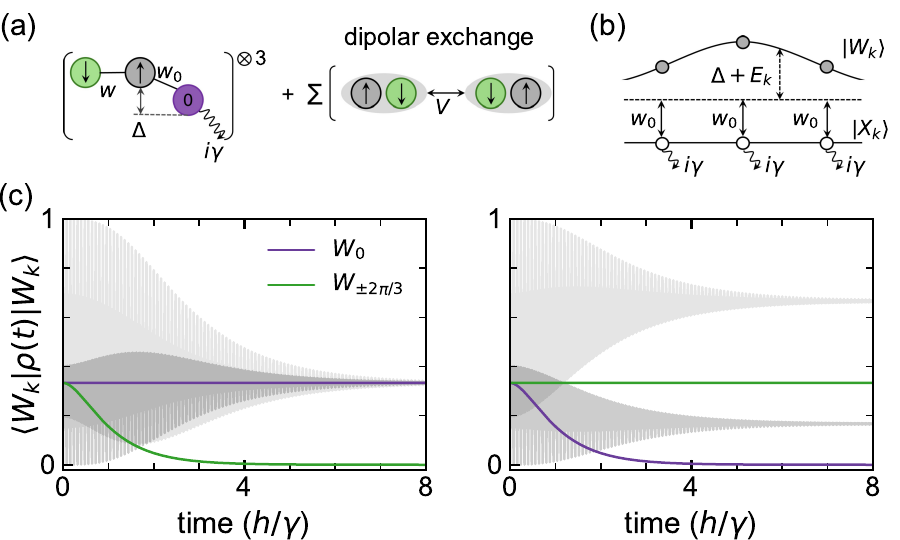}
	\caption{\textbf{Energy-selective distillation of 3-atom $W$ states.}
        \textbf{(a)}~Three atoms in a triangular spatial geometry with two internal Rydberg states encoded as $\ket{\uparrow}$ and $\ket{\downarrow}$. Within each atom, two Rydberg states are coupled with strength $w$, and the $\ket{\uparrow}$ state is coupled to the loss channel $\ket{0}$ (the loss rate is $\gamma$) with strength $w_0$ and detuning $\Delta$. Each two atoms undertake dipolar exchange with the strength $V$.
        \textbf{(b)}~Single-excitation subspace $\{\ket{W_k}\}$ with $k=-2\pi/3, 0, +2\pi/3$ from left to right, which is coupled to loss channels in the corresponding subspace $\{\ket{X_k}\}$ with strength $w_0$ and detuning $\Delta+E_k$. Here $E_k = 2V\cos{k}$. 
        \textbf{(c)}~Distillation results with $\Delta=V$ (left panel) and $\Delta=-2V$ (right panel) with an initial state in $\ket{\downarrow\uparrow\downarrow}$ by solving the full Lindblad master equation. The purple and green lines correspond to the temporal overlaps $\langle W_k|\rho(t)|W_k\rangle$ with the $\ket{W_k}$ states for $k=0$ and $k=\pm 2\pi/3$, respectively. The gray lines show the population in $\ket{\uparrow}$ for each atom, normalized to the temporal total population in $\ket{\uparrow}$ during the distillation process. Here the parameters used are: $w=0$, $w_0=0.2\gamma$, $V=50\gamma$.
}
\label{fig4}
\end{figure}

\textit{Three-body $W$ state distillation.--}
Motivated by the interaction-enabled selective pair loss studied above, we now theoretically explore the possibility to obtain specific target states via energy-dependent distillation in multi-atom systems in the $V \gg \gamma \gg w_0$ limit. As shown in Fig.~\pref{fig4}{a}, we consider the simplest 3-atom case in a triangular spatial geometry (thus with all-to-all connectivity). The dynamics are described by the master equation (\ref{eq1}), with the total Hamiltonian $H_{\rm tot} = V\sum_{\langle \alpha, \beta\rangle}(\ket{\uparrow}_\alpha\bra{\downarrow}\otimes\ket{\downarrow }_\beta\bra{\uparrow}+ {\rm H.c.}) + w_0\sum_\alpha(\ket{0}_\alpha\bra{\uparrow}+{\rm H.c.}) - \Delta\sum_\alpha\ket{0}_\alpha\bra{0}$ and the collapse operator $L_0=\sqrt{\gamma}\sum_\alpha\ket{g}_\alpha\bra{0}$ ($\alpha,\beta$ indicate the atom index). We focus on the single-excitation subspace expanded by $\ket{W_k}=\frac{1}{\sqrt{3}}(\ket{\uparrow\downarrow\downarrow} + e^{ik}\ket{\downarrow\uparrow\downarrow} + e^{i2k}\ket{\downarrow\downarrow\uparrow})$ with $k=0, \pm 2\pi/3$; see Fig.~\pref{fig4}{b}. The three Bloch states are respectively connected by $w_0$ to the corresponding states $\ket{X_k}=\frac{1}{\sqrt{3}}(\ket{0\downarrow\downarrow} + e^{ik}\ket{\downarrow0\downarrow} + e^{i2k}\ket{\downarrow\downarrow 0})$, each having a dissipation rate $\gamma$ in the loss subspace.
Due to the different eigenenergies $E_k=2V\cos(k)$ of the $\ket{W_k}$ states, the loss can be selectively triggered by letting the $w_0$ coupling be resonant with a specific $\ket{W_k}$ state, i.e., $\Delta=-E_k$. Figure \pref{fig4}{c} shows the simulation results with the atoms initialized in $\ket{\downarrow\uparrow\downarrow}$. By setting $\Delta=V$ ($\Delta=-2V$), the states $\ket{W_{\pm 2\pi/3}}$ ($\ket{W_0}$) are effectively distilled. Such an energy-selective distillation can easily be generalized to longer spin chains~\cite{SuppMats}, suggesting a $|k|$-selective tailoring of Bloch state dissipation.

\textit{Conclusion.--}
In summary, we have expanded the study of dissipation engineering in Rydberg systems~\cite{Signoles2014} by implementing laser-controlled dissipation and exploring its interplay with strong atom-atom interactions.
We observed interaction-induced shifts of exceptional points and enhanced decay rates. By carefully tuning the loss channel parameters, we realized selective spin protection via the two-body quantum Zeno effect, possibly extendable to dissipative chains for spin purification and distillation. Our results establish Rydberg arrays as a flexible platform for engineered open-system quantum simulation, allowing for studies of dissipative phase transitions~\cite{Lee2011,Perroni2023,Liu2024}, loss-induced collective synchronous phenomena and chirality~\cite{Li2025,Zhang2025}, and non-Hermitian many-body physics~\cite{Joshi2013,Butcher2022}. Moreover, our dissipation engineering scheme may be directly applied for quantum state preparations~\cite{Ramos2014,Jian2021} and extended realizations of dissipative spin models~\cite{Werner2005,Web2022}. For example, with the local addressing techniques, individual-site-resolved dissipation can be readily applied, to realize high-fidelity many-body state and study dissipation-enabled topology~\cite{Takata2018,Nie2021,Liang2022,Wetter2023}. 

We thank Tabor Electronics greatly for the use of an arbitrary waveform generator demo unit. This material is based upon work supported by the AFOSR MURI program under agreement number FA9550-22-1-0339 and by the National Science Foundation under grants No.~1945031 and No.~2438226.

\bibliographystyle{apsrev4-2}
\bibliography{lossRyd-edits}

\clearpage

\renewcommand{\thesection}{\Alph{section}}
\renewcommand{\thefigure}{S\arabic{figure}}
\renewcommand{\thetable}{S\Roman{table}}
\setcounter{figure}{0}
\renewcommand{\theequation}{S\arabic{equation}}
\renewcommand{\thepage}{S\arabic{page}}
\setcounter{equation}{0}
\setcounter{page}{1}

\newpage

\begin{widetext}
\appendix

\section{\large Supplemental Material for ``Collective dissipation engineering of \\ interacting Rydberg atoms''} 


\subsection{Renormalization of the experimental measurements}


As discussed in Ref.~\cite{Chen2024a}, the primary data we measure for the loss and $\ket{\uparrow}$ state population dynamics has a lower contrast as compared to the renormalized data presented in the main text and in the following sections. There are mainly two issues that set the upper and lower detection limits in our experiment. First, the inefficiency of the two-photon Rydberg excitation and the atom loss during trap release-and-recapture process lead to an average upper ``ceiling'' value $P_u(<1)$. Since the inefficient excitation leaves a fraction of atoms in ground state, it also contributes to the background lower limit $P_l$.
Second, the spontaneous decay of the Rydberg states (with typical lifetime of $40-50~\mu$s) provides another contribution to the lower value $P_l$.
This spontaneous decay of the Rydberg states is generally an order of magnitude lower than our laser-induced dissipation through the $\ket{5P_{1/2}}$ state. 
To note, the ``loss'' due to spontaneous decay does not appear as a time-dependent loss in our data (e.g., Fig.~1(c)), because it acts over a fixed time window and is independent of the application of our laser-induced loss pulses. Thus, we do not explicitly consider the contribution of spontaneous decay
(which is weaker and far less state-specific than our laser-induced loss) in our estimate of the loss rates $\gamma$. However, all data does account for such decay through the renormalization of measurements based on the independently measured $P_l$ values.

The data points in Fig.~1 are renormalized from $P_{\rm loss}^{\rm bare}$ to $P_{\rm loss}=(P_{\rm loss}^{\rm bare}-P_l)/(P_u-P_l)$ with $P_u = 0.93(1)$ and $P_l = 0.35(1)$. 
For the population loss after $1~\mu{\rm s}$ in Fig.~2, we renormalize the bare measured value $P_{\rm loss}^{\rm bare}$ to $P_{\rm loss}=(P_{\rm loss}^{\rm bare}-P_l)/(P_u-P_l)$ with $P_u = 0.92(1)$ and $P_l = 0.34(1)$. In Fig.~3, the population in $\ket{\uparrow}$ is resolved by measuring the total population $P_o^{\rm bare}$ in all other states: $\ket{\downarrow}$, $\ket{0}$ and $\ket{g}$. To accomplish this, we use a so-called ``bucket measurement'' approach by applying a coupling microwave for $\ket{\downarrow}\leftrightarrow\ket{0}$ when we de-excite atoms in $\ket{0}$ to the ground state for fluorescence detection, which effectively de-excites both $\ket{\downarrow}$ and $\ket{0}$. We first renormalize $P_o^{\rm bare}$ to $P_o = (P_o^{\rm bare}-P_l)/(P_u-P_l)$ with $P_u = 0.90(1)$ and $P_l=0.38(1)$. Then we determine the renormalized population in $\ket{\uparrow}$ simply as $P_\uparrow=1-P_o$. To note, the statistical variations of the renormalization factors are not accounted when performing this normalization, which will lead to additional uncertainties on the values of the renormalized population data.

\subsection{Non-interacting two-level dissipative systems across the exceptional point}

\begin{figure}[b]
	\includegraphics[width=0.95\textwidth]{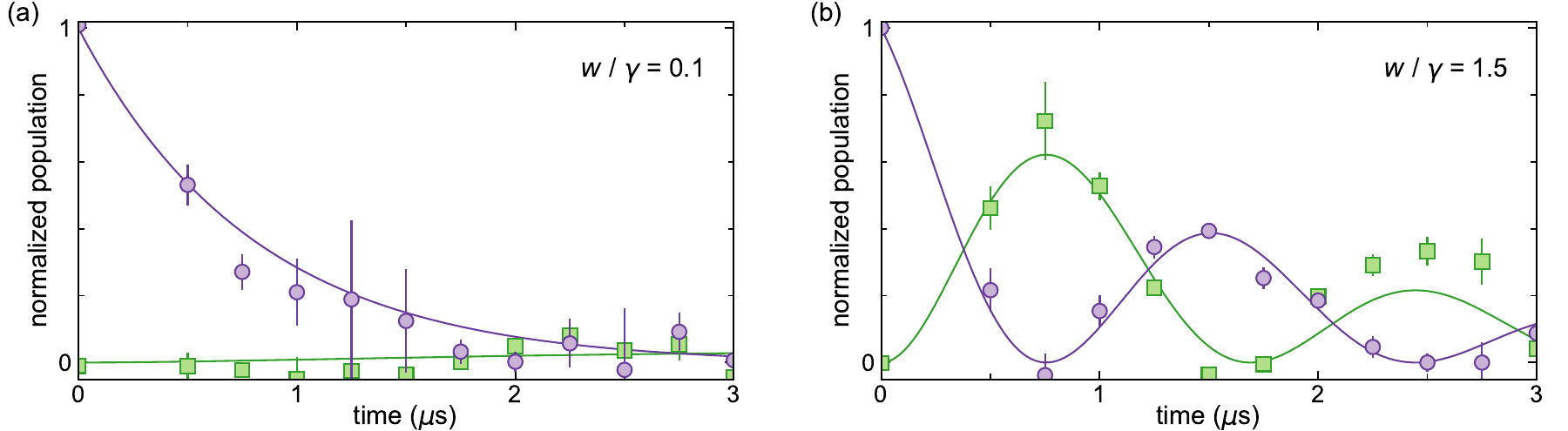}
	\caption{Time evolution of a non-interacting two-level dissipative system with different parameters: \textbf{(a)} $w/\gamma=0.1$ and \textbf{(b)} $w/\gamma=1.5$. We consider the system in Fig.~2(a) in the main text, with the atom initialized in $\ket{0}=\ket{42S_{1/2}}$ and the calibrated loss rate $\gamma/h=0.20(2)~{\rm MHz}$ for both (a) and (b). The normalized populations in $\ket{0}$ and $\ket{1}=\ket{42P_{3/2}}$ are shown in purple circles and green squares, respectively. All the solid lines are simulation results from the Lindblad master equation.
    Error bars are standard errors from multiple independent data sets.
	}
\label{figS1}
\end{figure}

We first benchmark our implementation of a two-level dissipative system without the consideration of the dipolar exchange interaction; see Fig.~2(a) in the main text. The dynamics are governed by the Lindblad master equation in the $\{\ket{1},\ket{0},\ket{g}\}$ space as
\begin{equation}\label{eqS1}
 \frac{\partial \rho}{\partial t} = -i[H_{\rm tot}, \rho] + L_0\rho L_0^\dagger -\frac{1}{2}\{L_0^\dagger L_0, \rho\} = \mathcal{L}\rho
\end{equation}
where $\rho$ denotes a $3\times 3$ density matrix and $\mathcal{L}$ is the Liouvillian superoperator with the Hamiltonian $H_{\rm tot} = w(\ket{0}\bra{1} + \ket{1}\bra{0})$ and the collapse operator $L_0=\sqrt{\gamma}\ket{0}\bra{g}$. By eliminating the quantum jumps, the dynamics within the $\{\ket{1},\ket{0}\}$ subspace can be effectively described by the non-Hermitian Hamiltonian
\begin{equation}
 H_{\rm eff} = \left(\begin{matrix}0 & w \\ w & -i\gamma/2\end{matrix}\right).
\end{equation}
It has been shown that the Liouvillian superoperator $\mathcal{L}$ and the above effective Hamiltonian share the same exceptional point at $w/\gamma=1/4$ \cite{Chen2021sc}. Note that this critical value corresponds to the ratio $w_c/\gamma=\sqrt{2}/4$ for the non-interacting two-atom case shown in Fig.~2(c) in the main text.

We observe different time evolution dynamics when scanning the $w/\gamma$ ratio across the exceptional point. In the small $w/\gamma$ limit, the system lies in $\mathcal{PT}$ symmetry-broken regime, and we observe a direct decay from $\ket{0}$ with essentially no population appearing in the upper $\ket{1}$ state; see Fig.~\pref{figS1}{a}. Since the $\ket{1}$ state is isolated, the system should avoid decay when initialized at $\ket{1}$ in the strong dissipation limit with large $\gamma$. However, when starting in $\ket{0}$ and increasing the $w/\gamma$ ratio, we observe oscillations between $\ket{0}$ and $\ket{1}$ that lead to a relatively slow decay of the whole system; see Fig.~\pref{figS1}{b}. 
This can be understood as the large $w$ coupling shifting the state $\ket{0}$ in energy, making it far-detuned from the dissipation channel and resulting in the quantum Zeno dynamics when starting from the $\ket{0}$ state~\cite{Facchi2002, Signoles2014}.

\subsection{Two-photon Raman transition between the two $42P$ states}

\begin{figure}[b]
	\includegraphics[width=0.7\textwidth]{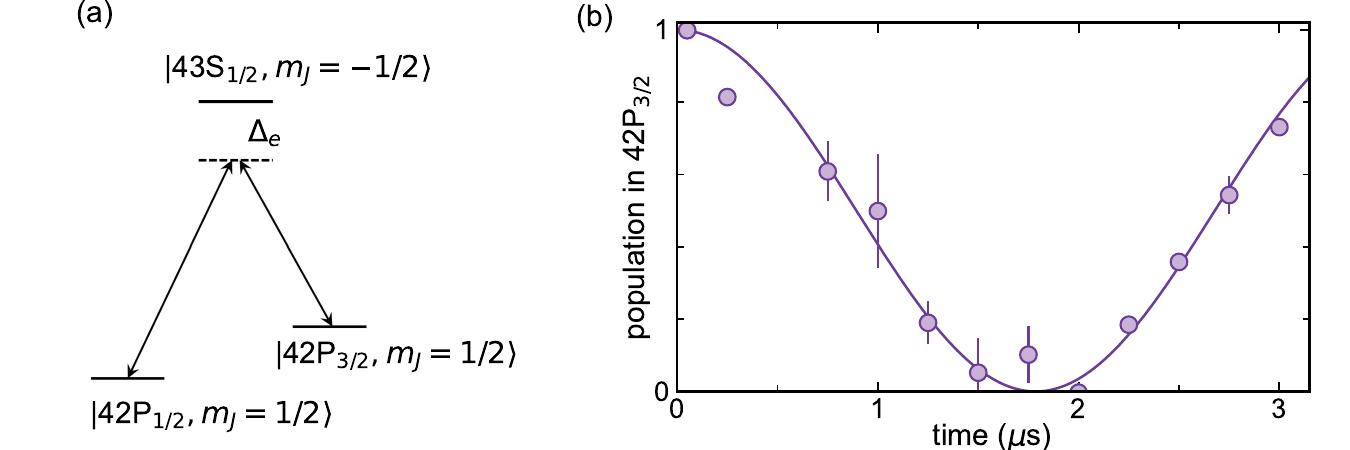}
	\caption{Two-photon Raman transition between two $42P$ states.
        \textbf{(a)}~Experimental scheme. Two sublevels $\ket{42P_{1/2},m_J=1/2}$ and $\ket{42P_{3/2},m_J=1/2}$ are coupled by two far-detuned microwave transitions via the upper $\ket{43S_{1/2},m_J=-1/2}$ state. Here $\Delta_e/h=16~{\rm MHz}$. 
        \textbf{(b)}~Measured Rabi oscillation between two sublevels. A cosine fit of the evolution of the population in $42P_{3/2}$ state yields the coupling rate of $w/h=0.15(2)~{\rm MHz}$.
}
\label{figS2}
\end{figure}

In Fig.~3(a) in the maintext, we encode a two-level, spin-1/2 system with $\ket{\uparrow}=\ket{42P_{3/2},m_J=1/2}$ and $\ket{\downarrow}=\ket{42P_{1/2},m_J=1/2}$. The two states are coupled via a two-photon microwave transition; see Fig.~\pref{figS2}{a}. The single photon detuning is set to $\Delta_e/h=16~{\rm MHz}$. We can tune the two single-photon Rabi rates (each much smaller than $\Delta_e$) to control the coupling strength $w$ between these two states. In our experiment, we first excite atoms into $\ket{\uparrow}=\ket{43P_{3/2},m_J=1/2}$, and then measure the time evolution, as shown in Fig.~\pref{figS2}{b}. A cosine fit yields the coupling rate $w/h=0.15(2)~{\rm MHz}$ that gives the $\pi$ time $t=1.7~\mu{\rm s}$. 

\subsection{Derivation of the reduced Hamiltonian (3) under specific parameter settings}

Here we show how to obtain the reduced effective Hamiltonian (3) in the main text. We still start from the Lindblad master equation (1) in the main text for the two-atom settings shown in Fig.~3(a), and eliminate the quantum jumps to the ground state $\ket{g}$. The full effective non-Hermitian Hamiltonian then reads
\begin{eqnarray}
 H_{\rm full} &=& \sum_{\alpha=A,B}(w\ket{\uparrow}_\alpha \bra{\downarrow} + w_0\ket{0}_\alpha\bra{\downarrow} + {\rm H.c.}) + \left( V_\uparrow\ket{\uparrow}_A\bra{0}\otimes\ket{0}_B\bra{\uparrow} + V_\downarrow \ket{\downarrow}_A\bra{0}\otimes\ket{0}_B\bra{\downarrow}+ {\rm H.c.}\right) \nonumber\\
 & - &(\Delta+i\gamma/2)\sum_{\alpha=A,B}\ket{0}_\alpha\bra{0}
\end{eqnarray}
Next, we move to the spin basis. Since all the singlets $\ket{-}_{rs}=(\ket{rs}-\ket{sr}/\sqrt{2}$ with $r,s \in \{\uparrow, \downarrow, 0\}$ are decoupled from all the (globally driven) microwave transitions, we write the above full Hamiltonian in the basis $\{\ket{\uparrow\uparrow},\ket{+}_{\uparrow\downarrow},\ket{\downarrow\downarrow},\ket{+}_{\uparrow 0},\ket{+}_{\downarrow 0},\ket{00}\}$ as
\begin{equation}\label{eqS4}
 H_{\rm spin} = \left(\begin{matrix}0 & w_c & 0 & 0 & 0 & 0 \\ w_c & 0 & w_c & w_0  & 0 & 0 \\ 0 & w_c & 0 & 0 & \sqrt{2} w_0 & 0 \\ 0 & w_0 & 0 & V_\uparrow - \Delta - i\gamma/2 & w & 0 \\ 0 & 0 & \sqrt{2} w_0 & w & V_{\downarrow} - \Delta -i\gamma/2 & \sqrt{2} w_0 \\ 0 & 0 & 0 & 0 & \sqrt{2} w_0 & -2\Delta-i\gamma \end{matrix}\right)
\end{equation}
with the collective two-body coupling $w_c=\sqrt{2}w$. 

We next inspect the couplings between $\ket{+}_{\downarrow 0}$ and other spin states. For $\ket{+}_{\downarrow 0}\leftrightarrow \ket{\downarrow\downarrow}$, we have the coupling strength $\sqrt{2}w_0$ along with a detuning $V_\downarrow - \Delta$; for $\ket{+}_{\downarrow 0}\leftrightarrow \ket{+}_{\uparrow 0}$, we have the coupling strength $w$ and the detuning $V_\uparrow-V_\downarrow$. To decouple the $\ket{+}_{\downarrow 0}$ state, we can let
\begin{eqnarray}
 |V_\downarrow - \Delta|  &\gg w_0, \nonumber\\
 |V_\uparrow-V_\downarrow|  &\gg w.
\end{eqnarray}
Under these conditions, the $\ket{00}$ state can also be neglected as it only couples with the $\ket{+}_{\downarrow 0}$ state. We finally arrive at the reduced Hamiltonian (3) within the $\{\ket{\uparrow\uparrow},\ket{+}_{\uparrow\downarrow},\ket{\downarrow\downarrow},\ket{+}_{\uparrow 0}\}$ basis
\begin{equation}\label{eqS5}
 H_{\rm eff} = \left(\begin{matrix}0 & w_c & 0 & 0 \\ w_c & 0 & w_c & w_0 \\ 0 & w_c & 0 & 0 \\ 0 & w_0 & 0 & V_\uparrow - \Delta - i\gamma/2\end{matrix}\right)
\end{equation}
which describes the scheme shown in Fig.~3(b) in the main text.

We provide further evidence for the above Hamiltonian reduction conditions by comparing the eigenenergy spectra for the full spin Hamiltonian (\ref{eqS4}) and the reduced effective Hamiltonian (\ref{eqS5}). As shown in Fig.~\pref{figS3}{b}, with $\Delta=V_\uparrow = 500\gamma$, $V_\downarrow=V_\uparrow/2$, and $w=\gamma$, the imaginary parts of 4 eigenenergies approximately approach those for the reduced case in Fig.~\pref{figS3}{a}. Note the different axis scales here from that in Fig.~3(c) in the maintext. 

\begin{figure}[h]
	\includegraphics[width=0.95\textwidth]{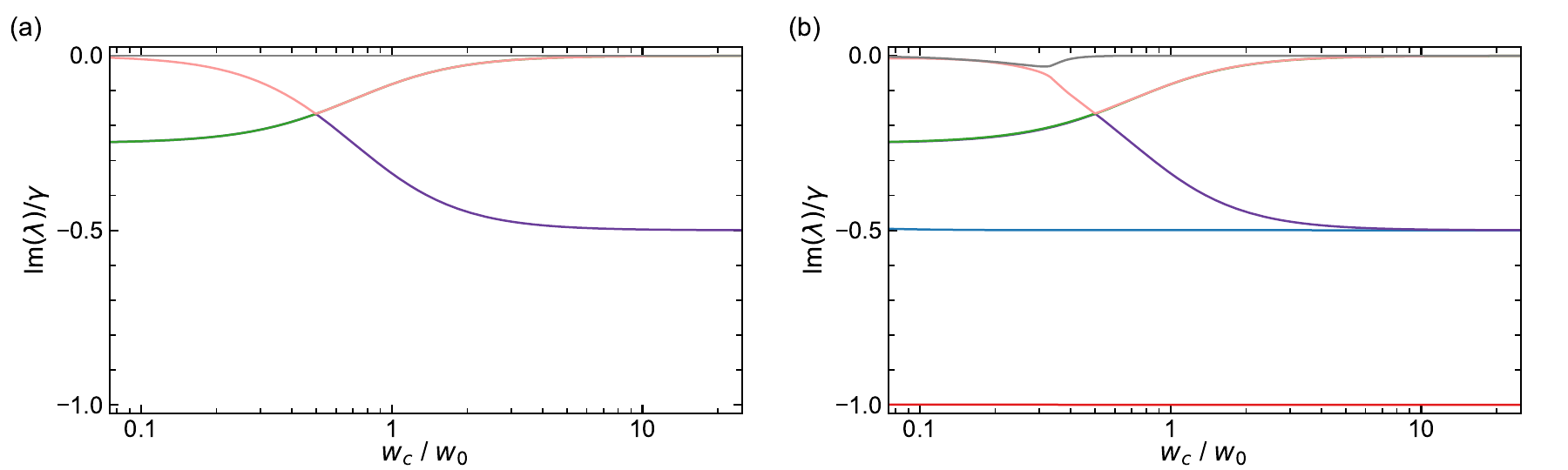}
	\caption{Comaprison of the eigenenergy spectra for the full spin Hamiltonian (\ref{eqS4}) and the reduced effective Hamiltonian (\ref{eqS5}). 
        \textbf{(a)}~Imaginary parts of the eigenenergy values $\lambda$ versus $w_c/w_0$ ratios for Hamiltonian (\ref{eqS5}) with $\Delta=V_\uparrow$.  
        \textbf{(b)}~Imaginary parts of the eigenenergy values $\lambda$ versus $w_c/w_0$ ratios for the full spin Hamiltonian (\ref{eqS4}) with $\Delta=V_\uparrow=500\gamma$, $V_\downarrow=V_\uparrow/2$.
}
\label{figS3}
\end{figure}

\subsection{Comparison of single-atom and atom-pair dynamics in the $\Delta\gg w_0$ limit}

\begin{figure}[]
	\includegraphics[width=0.95\textwidth]{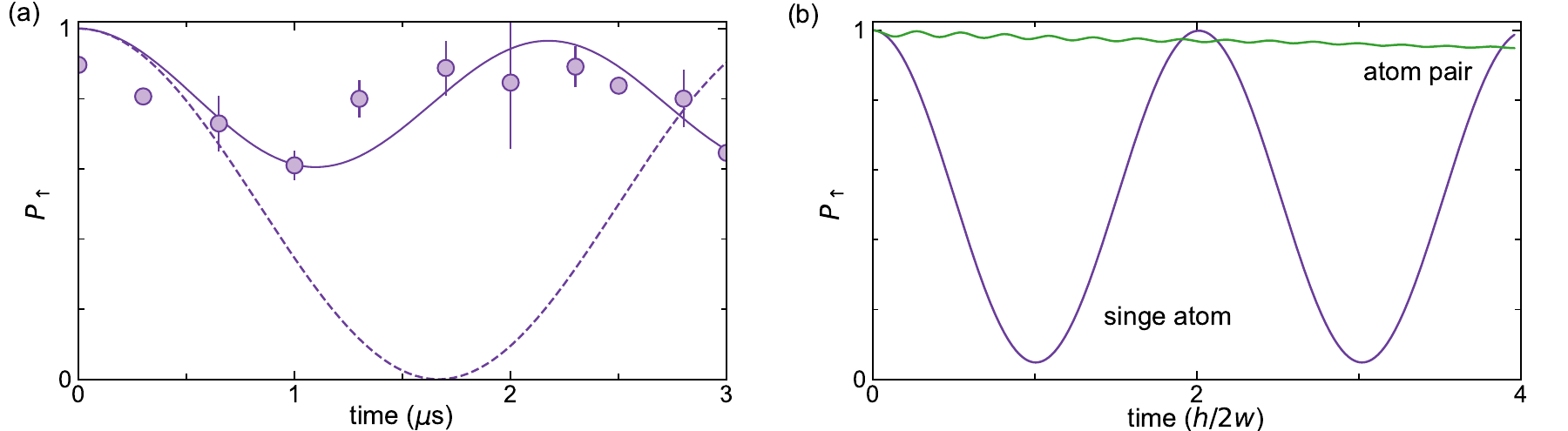}
	\caption{
    \textbf{(a)}~Experimentally measured time evolution of $P_\uparrow$ for single atoms with $\Delta/w_0=3.2$ ($\Delta/h=4.0(1)~{\rm MHz}$ and $w_0/h=1.25(2)~{\rm MHz}$, other parameters are the same as those in Fig.~3(e) in the main text). The solid line indicates the simulation result from the Lindblad master equation, while the dashed line gives the Rabi oscillation for $\ket{\uparrow}\leftrightarrow\ket{\downarrow}$ in $\Delta/w_0\to\infty$ limit.
    \textbf{(b)}~Single-atom (purple) and atom-pair (green) dynamics in the extreme two-body Zeno limit for the scheme in Fig.~3 of the main text. Here, we show the simulation results by solving the Lindblad master equation with $V_\uparrow=\Delta=500\gamma, V_\downarrow=V_\uparrow/2, w_0=15\gamma, w=\gamma$. The initial state assumes a pair of atoms in $\ket{\uparrow\uparrow}$.
}
\label{figS4}
\end{figure}

In the two-body Zeno limit for the settings in Fig.~3 in the main text, we expect entirely different dynamics for single atom and atom pairs, as the Zeno effect is triggered by the dipolar exchange interactions. In the left panel of Fig.~3(e), we experimentally observe the nearly frozen dynamics for atom pairs initialized in the $\ket{\uparrow\uparrow}$ configuration when $\Delta/w_0=3.2$ and $w/w_0=0.12$. 
For this not-very-large $\Delta/w_0$ ratio, the single atom dynamics do not follow a good two-level Rabi oscillation, as the transition $\ket{\downarrow}\leftrightarrow\ket{0}$ is still effectively coupled. As shown in Fig.~\pref{figS4}{a}, while the measured dynamics are different from the nearly frozen atom-pairs, they are still very different from the idealized dynamics of the $\Delta/w_0\to\infty$ limit.

By simply increasing the $\Delta$ value in the $V_\uparrow=\Delta\gg w_0 \gg w$ regime, the single-atom coupling $\ket{\uparrow}\leftrightarrow\ket{\downarrow}$ is isolated from the loss channel, leading to an almost perfect Rabi oscillation as shown in Fig.~\pref{figS4}{b}. For atom pairs, however, since the interaction compensates the large detuning and activates the loss channel, they will suffer from a strong dissipation in fact experience an effectively slowed down or inhibited decay due to the (pair state) quantum Zeno effect.

\subsection{Dynamics for a dissipative chain}

\begin{figure}[b]
	\includegraphics[width=0.5\columnwidth]{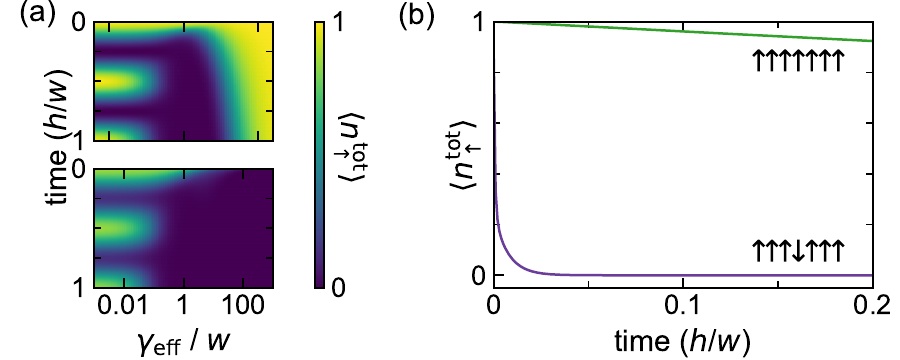}
	\caption{Dynamics of a chain with imaginary spin exchanges.
        \textbf{(a)}~Time evolution of the total population fraction in $\ket{\uparrow}$, $\langle n_\uparrow^{\rm tot}\rangle$, for different dissipation rates $\gamma_{\rm eff}$ with the initial state: (upper panel) $\ket{\uparrow\uparrow\uparrow\uparrow\uparrow\uparrow\uparrow}$ and (lower panel) $\ket{\uparrow\uparrow\uparrow\downarrow\uparrow\uparrow\uparrow}$.  
        \textbf{(b)}~Short time dynamics of $\langle n_\uparrow^{\rm tot}\rangle$ for the two different spin configurations in (a) with $\gamma_{\rm eff}/w=100$. The results are obtained with the effective non-Hermitian Hamiltonian (\ref{eq5}).
}
\label{figS5}
\end{figure}

We can generalize the effective two-body Hamiltonian (4) in the main text for a long spin chain of $N$ atoms under periodic boundary conditions. Restricting the interactions to nearest-neighbors, the Hamiltonian takes the form
\begin{equation}\label{eq5}
 H_{\rm chain} = w\sum_j\sigma_j^x -\frac{i\gamma_{\rm eff}}{4}\sum_j(\sigma_j^x\sigma_{j+1}^x + \sigma_j^y\sigma_{j+1}^{y} + I_jI_{j+1} - \sigma_j^z\sigma_{j+1}^z) \ ,
\end{equation}
where the second term contains an imaginary spin exchange term that avoids population in configurations with neighboring spins anti-aligned. In other words, when two adjacent atoms are in different spins, they experience rapid decay. 

In Fig.~\ref{figS5}, this mechanism leads to strikingly different many-body dynamics for different spin configurations. For a fully polarized state $\ket{\uparrow\uparrow\dots\uparrow}$, the quantum Zeno effect in the large $\gamma_{\rm eff}$ limit dynamically protects the system from spin flips and stabilizes it in the initial configuration. In contrast, introducing a single spin flip at $t=0$ entirely changes the evolution. Large $\gamma_{\rm eff}$ now accelerates the decay of the whole chain, as evidenced by the rapid drop of the average $\ket{\uparrow}$ fraction $\langle n_\uparrow^{\rm tot}\rangle$ with $n_\uparrow^{\rm tot} = \frac{1}{N}\sum_j\ket{\uparrow}_j\bra{\uparrow}$ in Fig.~\pref{figS5}{b}. This selective protection/decay mechanism can, in principle, be exploited as a dissipative purification protocol, by filtering out unwanted configurations containing spin-down defects while preserving ferromagnetically ordered states or possibly conserving other desired correlated subspaces.

\subsection{$|k|$-selective tailoring of the Bloch states in long spin chains}

\begin{figure}[b]
	\includegraphics[width=0.9\columnwidth]{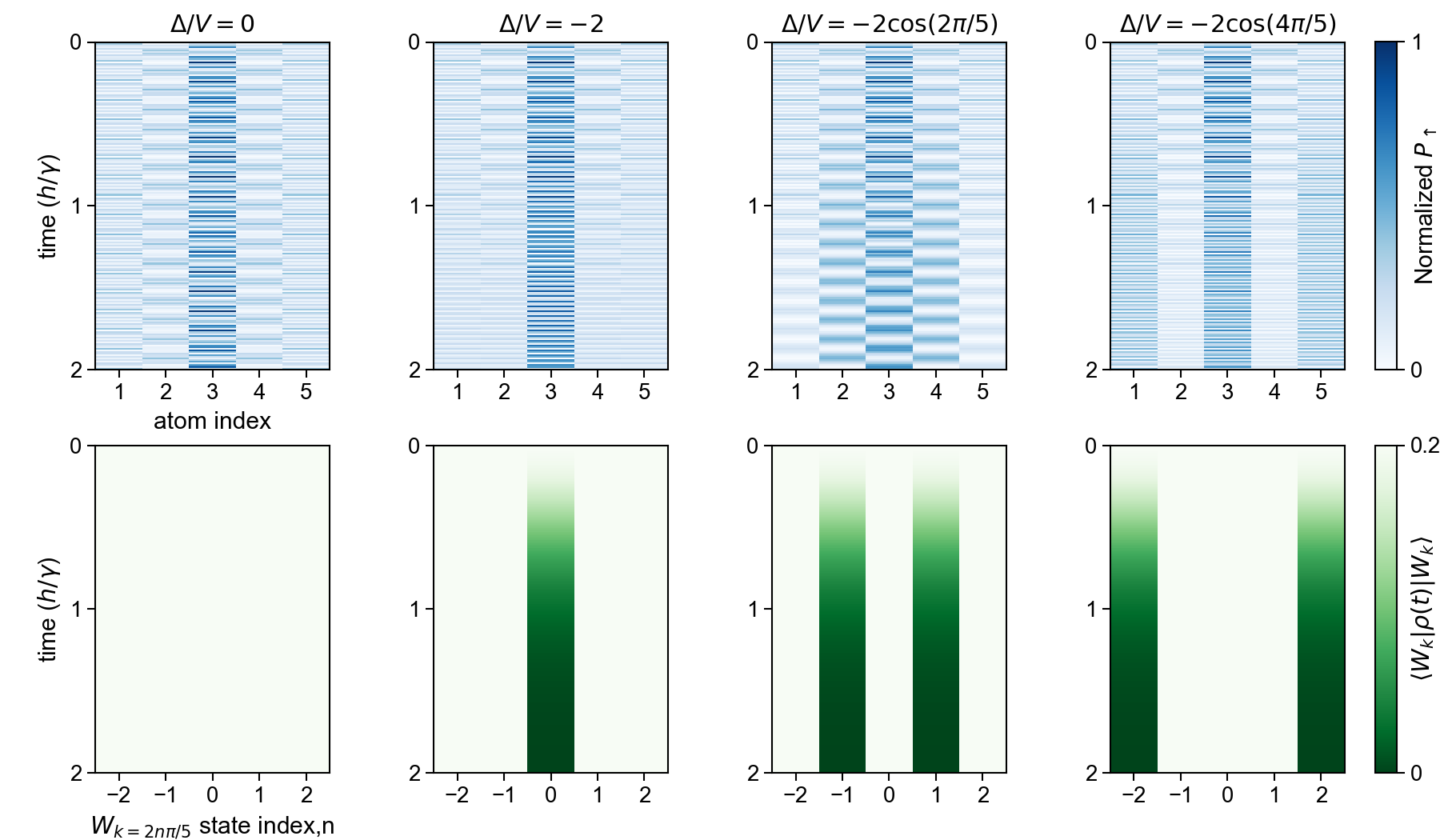}
	\caption{Time evolutions of $|k|$-selective loss in a 5-atom spin chain under periodic boundary conditions with the initial state $\ket{\downarrow\downarrow\uparrow\downarrow\downarrow}$.
        \textbf{(Upper panel)}~Time evolution of the normalized population in $\ket{\uparrow}$ for each atom, under different detuning choices. From left to right, the detuning $\Delta=\{0, -2V, 2V\cos{(2\pi/5)}, 2V\cos{(4\pi/5)}\}$, respectively.
        \textbf{(Lower panel)}~Time evolution of the overlap $\langle W_k|\rho(t)\ket{W_k}$ between the temporal state and the single-excitation Bloch states $\ket{W_{k=2n\pi/5}}$. We can see clear selective damping of specific $W_k$ states by matching to resonant loss coupling channels (see Fig.~4 in the main text). 
        Here the results are from full simulations with the Lindblad master equation, with the parameters $w=0$, $w_0 = 0.3\gamma$, $V=30\gamma$. 
}
\label{figS6}
\end{figure}

Here we extend the energy-selective $W$ state distillation to longer multiple-atom chains. Let us take a 5-atom chain under periodic boundary conditions as an example, and only consider nearest-neighbor interactions for simplicity. The level scheme is shown in Fig.~4(a) of the main text. The dynamics are determined by the Lindblad master equation (1) in the main text, but with the Hamiltonian
\begin{equation}
 H_{\rm tot} = V\sum_{\langle \alpha, \beta\rangle}(\ket{\uparrow}_\alpha\bra{\downarrow}\otimes\ket{\downarrow}_\beta\bra{\uparrow} + {\rm H.c.}) + w_0\sum_\alpha(\ket{0}_\alpha\bra{\uparrow}+{\rm H.c.}) - \Delta\sum_\alpha\ket{0}_\alpha\bra{0}
\end{equation}
and the collapse operator $L_0=\sqrt{\gamma}\sum_\alpha\ket{g}_\alpha\bra{0}$, with $\alpha, \beta$ the atom index and $\langle \alpha, \beta\rangle$ the nearest-neighbored two atoms. We still consider the single-excitation subspace spanned by the Bloch eigenstates
\begin{equation}
 \ket{W_{k}} = \frac{1}{\sqrt{5}}(\ket{\uparrow\downarrow\downarrow\downarrow\downarrow} + e^{ik}\ket{\downarrow\uparrow\downarrow\downarrow\downarrow}+e^{i2k}\ket{\downarrow\downarrow\uparrow\downarrow\downarrow}+e^{i3k}\ket{\downarrow\downarrow\downarrow\uparrow\downarrow} + e^{i4k}\ket{\downarrow\downarrow\downarrow\downarrow\uparrow})
\end{equation}
with $k=2n\pi/5$, $n\in [-2, 2]$. Each $\ket{W_k}$ state is coupled to the lossy subspace spanned by 
\begin{equation}
\ket{X_k} = \frac{1}{\sqrt{5}}(\ket{0\downarrow\downarrow\downarrow\downarrow} + e^{ik}\ket{\downarrow 0\downarrow\downarrow\downarrow}+e^{i2k}\ket{\downarrow\downarrow 0\downarrow\downarrow}+e^{i3k}\ket{\downarrow\downarrow\downarrow 0\downarrow} + e^{i4k}\ket{\downarrow\downarrow\downarrow\downarrow 0})
\end{equation}
with a loss rate of $\gamma$ (see Fig.~3(b) in the main text). Since the eigenstates $\ket{W_k}$ have different energies $E_k=2V\cos{(k)}$, we can selectively let specific $|k|$ states damp by varying the detuning of the $w_0$ coupling $\ket{0}\leftrightarrow\ket{\uparrow}$. 

Figure \ref{figS6} shows the simulation results for such a 5-atom chain. Starting from $\ket{\downarrow\downarrow\uparrow\downarrow\downarrow} = \sum_n e^{-i4\pi/5}\ket{W_{k=2n\pi/5}}/\sqrt{5}$, with zero detuning of $\Delta$ (from the single-atom resonance condition), we see the oscillation dynamics of the spin-up excitation within the 5 atoms, and the overlaps between the temporal state and the 5 eigenstates are always 0.2, indicating no loss happens as all $\ket{W_k}\leftrightarrow\ket{X_k}$ transitions are far detuned from available many-atom states. However, by simply letting $\Delta=-E_k=-2V\cos{(2n\pi/5)}$, we can trigger the loss of $\ket{W_{k=\pm 2n\pi/5}}$ states, i.e., the stationary oscillations of $P_\uparrow$ shows different behaviors from that for $\Delta=0$. We can also see clear $|k|$-selective distillation by plotting the overlaps of the temporal state to the $W_k$ states in Fig.~\ref{figS6}. This scenario with a larger number of atoms is quite a bit richer than the three-atom case explored in the main text, as here one can retain non-equilibrium dynamics even after distilling some subset of the available many-atom Bloch states.

This scheme can be generalized to longer atomic chains as well as higher-dimensional Rydberg atom arrays. To note, when considering the single excitation sector (cases of only a single $\ket{\uparrow}$ excitation), the dynamics can effectively be captured by a single-particle formalism with a non-Hermitian Hamiltonian (by introducing an auxiliary, biased set of states experience loss analogous to the depiction of Fig.~4(b) in the main text).

\clearpage
\end{widetext}

\end{document}